\documentclass[a4paper,10pt,twoside]{cpc-hepnp}

\usepackage{multicol}
\usepackage{graphicx}
\usepackage{booktabs}
\usepackage{amssymb,bm,mathrsfs,bbm,amscd}
\usepackage[tbtags]{amsmath}
\usepackage{lastpage}
\usepackage{mdwlist}

\usepackage{graphicx}
\usepackage{xspace}
\usepackage{amsfonts}
\usepackage{srcltx}
\usepackage{hyperref}
\usepackage{psfrag}

\renewcommand{\epsilon}{\varepsilon}

\begin{document}

\fancyhead[co]{\footnotesize V.V. Anashin et al., Recent results from
  the KEDR Detector} \title{Recent results from the KEDR Detector
  \thanks{Partially supported by the Russian Foundation for Basic
    Research, Grant 04-02-16712, 07-02-00816, 07-02-01162 and RF
    Presidential Grant for Sc. Sch. NSh-5655.2008.2 } }

\author{
  V.V. Anashin$^1$, V.M. Aulchenko$^{1,2}$, E.M. Baldin$^{1,2}$, A.K. Barladyan$^1$, A.Yu. Barnyakov$^1$, M.Yu. Barnyakov$^1$, \\
  S.E. Baru$^{1,2}$, I.V. Bedny$^1$, O.L. Beloborodova$^{1,2}$, A.E. Blinov$^{1;1)}$,\email{A.E.Blinov@inp.nsk.su} V.E. Blinov$^{1,3}$, A.V. Bobrov$^1$, \\
  V.S. Bobrovnikov$^1$, A.V. Bogomyagkov$^{1,2}$, A.E. Bondar$^{1,2}$, A.R. Buzykaev$^1$, S.I. Eidelman$^{1,2}$, \\
  Yu.M. Glukhovchenko$^1$, V.V. Gulevich$^1$, D.V. Gusev$^1$, S.E. Karnaev$^1$, S.V. Karpov$^1$, T.A. Kharlamova$^{1,2}$, \\
  V.A. Kiselev$^1$, S.A. Kononov$^{1,2}$, K.Yu. Kotov$^1$, E.A. Kravchenko$^{1,2}$, V.F. Kulikov$^{1,2}$, G.Ya. Kurkin$^{1,3}$, \\
  E.A. Kuper$^{1,2}$, E.B. Levichev$^{1,3}$, D.A. Maksimov$^1$, V.M. Malyshev$^1$, A.L. Maslennikov$^1$, A.S. Medvedko$^{1,2}$, \\
  O.I. Meshkov$^{1,2}$, S.I. Mishnev$^1$, I.I. Morozov$^{1,2}$, N.Yu. Muchnoi$^{1,2}$, V.V. Neufeld$^1$, S.A. Nikitin$^1$, \\
  I.B. Nikolaev$^{1,2}$, I.N. Okunev$^1$, A.P. Onuchin$^{1,3}$, S.B. Oreshkin$^1$, I.O. Orlov$^{1,2}$, A.A. Osipov$^1$, \\
  S.V. Peleganchuk$^1$, S.G. Pivovarov$^{1,3}$, P.A. Piminov$^1$, V.V. Petrov$^1$, A.O. Poluektov$^1$, I.N. Popkov$^1$, V.G. Prisekin$^1$, \\
  A.A. Ruban$^1$, V.K. Sandyrev$^1$, G.A. Savinov$^1$, A.G. Shamov$^1$, D.N. Shatilov$^1$, B.A. Shwartz$^{1,2}$, E.A. Simonov$^1$, \\
  S.V. Sinyatkin$^1$, Yu.I. Skovpen$^{1,2}$, A.N. Skrinsky$^1$, V.V. Smaluk$^{1,2}$, A.V.~Sokolov$^1$, A.M. Sukharev$^1$, \\
  E.V. Starostina$^{1,2}$, A.A. Talyshev$^{1,2}$, V.A. Tayursky$^1$, V.I. Telnov$^{1,2}$, Yu.A. Tikhonov$^{1,2}$, K.Yu. Todyshev$^{1,2}$, \\
  G.M. Tumaikin$^1$, Yu.V. Usov$^1$, A.I. Vorobiov$^1$, A.N. Yushkov$^1$, V.N. Zhilich$^1$, V.V.Zhulanov$^{1,2}$,  A.N. Zhuravlev$^{1,2}$ \\
} \maketitle

\address{
  1~(Budker Institute of Nuclear Physics, 11, Lavrentiev prospect,
  Novosibirsk, 630090, Russia)\\
  2~(Novosibirsk State University, 2, Pirogova street, Novosibirsk, 630090, Russia)\\
  3~(Novosibirsk State Technical University, 20, Karl Marx prospect,
  Novosibirsk, 630092, Russia) }

  \begin{abstract}
    We report results of experiments performed with the KEDR detector
    at the VEPP-4M $e^+e^-$ collider.  They include precise
    measurement of the $D^0$ and $D^{\pm}$ meson masses, determination
    of the $\psi(3770)$ resonance parameters, and a search for narrow
    resonances in $e^+e^-$ annihilation at center-of-mass energies
    between 1.85 and 3.1 GeV.

  \end{abstract}


\begin{multicols}{2}

\section{Introduction}
\label{sec:intro}

The paper reports results of three analyses performed with the KEDR
detector at the VEPP-4M collider (BINP, Novosibirsk): a measurement of
$D^0$ and $D^{\pm}$ meson masses, a measurement of $\psi(3770)$ mass
and widths, and a search for narrow resonances in $e^+e^-$
annihilation between 1.85 and 3.1 GeV.

\section{VEPP-4M collider and KEDR detector}
\label{sec:VEPPKEDR}

The VEPP-4M collider~\cite{Anashin:1998sj} can operate in the wide
range of beam energy from 2E=~2 to 12 GeV.  The peak luminosity in the
\(J/\psi\) energy region is
about~\(2\times10^{30}\,\text{cm}^{-2}\text{s}^{-1}\).
One of the main features of \\
\begin{center}
  \includegraphics[width=\columnwidth]{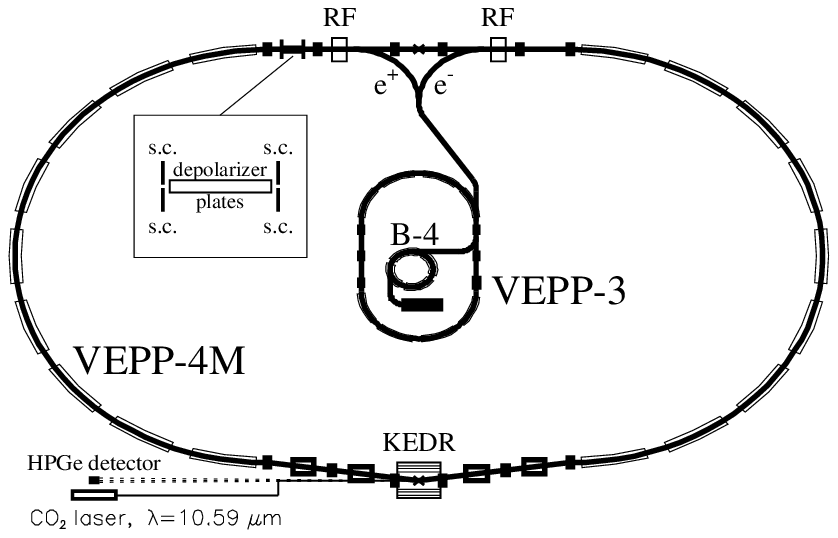}
  \figcaption{\label{fig:vepp4m}
       VEPP-4M/KEDR complex with the resonant depolarization and
    the infrared light Compton backscattering facilities.}
\end{center}
the VEPP-4M is its capability to precisely measure the beam energy
using two techniques~\cite{Blinov:2009}: resonant depolarization and
infrared light Compton backscattering.

The accuracy of VEPP-4M energy measurement with the resonant
depolarization reaches \(10^{-6}\).  Such measurement requires
dedicated calibration runs without data taking.  The accuracy of the
beam energy determination for the accumulated data sample is dominated
by the interpolation between the successive depolarization runs and
equals about \(6\cdot10^{-6}\) ($\simeq$10~keV) in the \(J/\psi\)
region~\cite{Aulchenko:2003qq}.

A new technique developed at the BESSY-I and BESSY-II synchrotron
radiation sources~\cite{Klein:1997wq,Klein:2002ky} was adopted for
VEPP-4M in 2005.  It employs the infrared light Compton backscattering
and has a worse precision compared to the resonant
depolarization(50$\div$70~keV in the \(J/\psi\) region), but unlike
the latter it can be used during data taking~\cite{Blinov:2009}.

The KEDR detector~\cite{Anashin:2002uj} includes the vertex detector,
the drift chamber, the scintillation time-of-flight counters, the
aerogel Cherenkov counters, the barrel liquid krypton calorimeter, the
endcap CsI calorimeter, and the muon system embedded in the yoke of a
superconducting coil generating a field of 0.6 T.  The detector also
includes a high-resolution tagging system for studies of two-photon
processes.  The on-line luminosity measurement is performed with two
independent single bremsstrahlung monitors.

\begin{center}
  \includegraphics[width=\columnwidth]{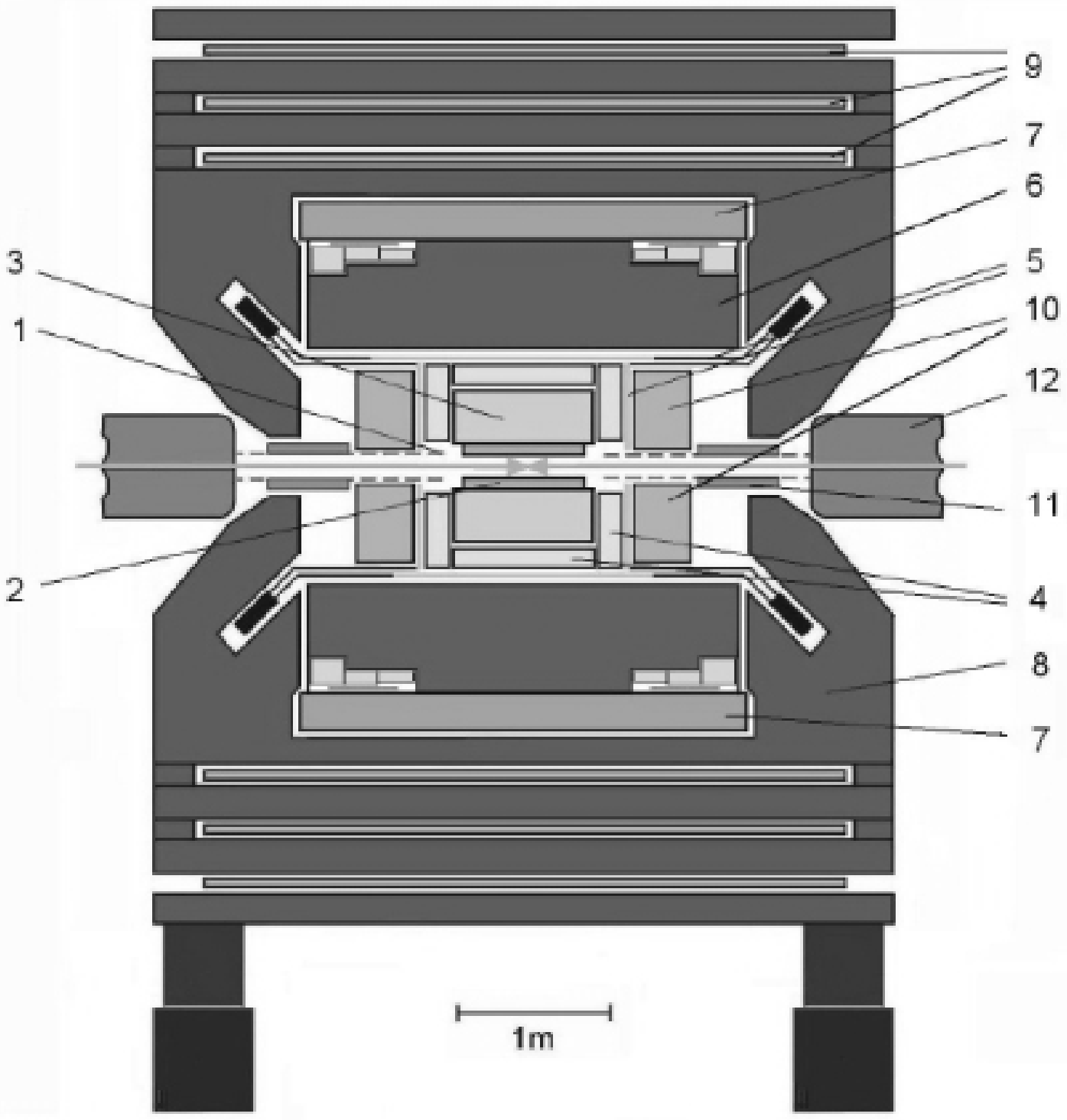}
  \figcaption{\label{fig3}KEDR detector. 1-vacuum chamber, 2-vertex
    detector, 3-drift chamber, 4-threshold aerogel counters,
    5-ToF-counters, 6-liquid krypton calorimeter, 7-superconducting
    coil (0.6 T), 8-magnet yoke, 9-muon tubes, 10-CsI-calorimeter,
    11-compensation solenoid, 12-VEPP-4M quadrupole.  }
\end{center}

\section{$D^0$ and $D^{\pm}$ meson masses}
\label{sec:Dmass}

Neutral and charged $D$ mesons are the ground states in the family of
open charm mesons.  Measurement of their masses provides a mass scale
for the heavier excited states.  In addition, a precise measurement of
the $D^0$ meson mass could help to understand the nature of the narrow
$X(3872)$
state~\cite{Choi:2003ue,Acosta:2003zx,Abazov:2004kp,Aubert:2004ns},
which, according to some models, is a bound state of $D^0$ and
$D^{*0}$~\cite{Swanson:2003tb} mesons and has a mass very close to the
sum of the $D^0$ and $D^{*0}$ meson masses.

Measurement of $D$ meson masses is performed using the near-threshold
$e^+e^-\to D\overline{D}$ production at the $\psi(3770)$ resonance
with the full reconstruction of one of the $D$ mesons.  Neutral and
charged $D$ mesons are reconstructed in the $K^-\pi^+$ and
$K^-\pi^+\pi^+$ final states, respectively (charge-conjugate states
are implied throughout this paper).  The analysis uses a data sample
of 0.9 pb$^{-1}$.

The mass of the $D$ meson is calculated as
\begin{equation}
    M_{\rm bc}\simeq\sqrt{E_{\rm beam}^2-
               \left(\sum\limits_i \vec{p}_i\right)^2},
\end{equation}
(so-called {\it beam-constrained mass}), where $E_{\rm beam}$ is the
average energy of colliding beams and $\vec{p}_i$ are the momenta of
the $D$ decay products.  Since $E_{\rm beam}$ is measured by resonant
depolarization technique as described in Sec.~\ref{sec:VEPPKEDR}, its
contribution to the $D$ mass error is negligible.

Apart of the $M_{\rm bc}$, $D$ decays are effectively selected by
energy difference
\begin{equation}
    \Delta E=\sum\limits_i \sqrt{M^2_i+p^2_i}-E_{\rm beam}\,
\end{equation}
where $M_i$ and $p_i$ are the masses and momenta of the $D$ decay
products.  The signal events satisfy a $\Delta E\simeq 0$ condition.
Therefore, the scale of momenta is tuned in order to get
$\langle\Delta E\rangle$ of $D$ decays consistent with zero.
Furthermore, while calculating $M_{\rm bc}$, we employ a kinematic fit
with the $\Delta E=0$ constraint.  It results in a certain improvement
of the $M_{\rm bc}$ resolution and significantly reduces a
contribution of the remaining momentum scale error to the measured $D$
mass.

In order to measure the $D$ mass most efficiently, the unbinned
maximum likelihood fit procedure is used.  In the case of $D^+$ meson
analysis, the likelihood function depends on two already defined
variables: $M_{\rm bc}$ and $\Delta E$.  The $D^0$ meson analysis
employs one more variable: the difference of the absolute values of
momenta of $D^0$ decay products in the CM frame $\Delta|p|$.  We use
the fact that the $M_{\rm bc}$ resolution strongly depends on decay
kinematics --- it is about three times better if daughter particles
move transversely to the $D^0$ direction (small $\Delta|p|$), than if
they move along this direction (large $\Delta|p|$).  Thus, the
$\Delta|p|$ variable estimates the $M_{\rm bc}$ resolution on the
event-by-event basis, improving the overall statistical accuracy of
the measurement.

A fit of the event density is performed with $D$ mass as one of the
parameters in a relatively wide region around $M_{\rm bc}=M_D$ and
$\Delta E=0$ (specifically, $M_{\rm bc}>1700$~MeV, $|\Delta
E|<300$~MeV), with the background contribution taken into account.
The likelihood function has the form:
\begin{equation}
 -2\log\mathcal{L}({\bf\alpha})=-2\sum\limits_{i=0}^{N}\log p({\bf v}_i| {\bf\alpha})+
  2N\log\int\!\!p({\bf v}| {\bf\alpha}) d{\bf v}\,
\end{equation}
where ${\bf v}=(M_{\rm bc}, \Delta E, \Delta |p|)$ are the variables
that characterize one event, $p({\bf v}| {\bf \alpha})$ is the
probability distribution function (PDF) of these variables depending
on the fit parameters ${\bf\alpha}=(M_D, \langle\Delta E\rangle,
b_{uds}, b_{DD})$:
\begin{equation}
 p({\bf v}|{\bf\alpha}) = p_{sig}({\bf v}|M_D, \langle\Delta E\rangle)+
   b_{uds}p_{uds}({\bf v})+
   b_{DD}p_{DD}({\bf v})\,.
 \label{exp_pdf}
\end{equation}
Here $p_{sig}$ is the PDF of the signal events which depends on $M_D$
and $\langle\Delta E\rangle$ (the central value of the $\Delta E$
distribution), $p_{uds}$ is the PDF for the background process
$e^+e^-\to q\overline{q}$ ($q=u,d,s$), and $p_{DD}$ is the PDF for the
background from $e^+e^-\to D\overline{D}$ decays with $D$ decaying to
all modes other than the signal one. $b_{uds}$ and $b_{DD}$ are
relative magnitudes of the background terms.  The shapes of the
$p_{sig}$, $p_{uds}$ and $p_{DD}$ distributions are obtained from the
MC simulation.

The simulation of signal events is performed with the MC generator for
$e^+e^-\to D\overline{D}$ decays where $D$-meson decays are simulated
by the {\tt JETSET 7.4} package~\cite{Sjostrand:1986hx}.  The
radiative corrections are taken into account in both initial (the {\tt
  RADCOR} package~\cite{radcor} with Kuraev-Fadin
model~\cite{KuraevFadin}), and final states (the {\tt PHOTOS}
package~\cite{photos}).  The ISR corrections use the $e^+e^-\to
\psi(3770)\to D\overline{D}$ cross section according to a Breit-Wigner
amplitude with $M=3771$ MeV and $\Gamma=23$ MeV~\cite{PDG2006}.  The
backgrounds from the continuum $e^+e^-\to q\bar{q}$ process and from
$e^+e^-\to D\overline{D}$ decays are simulated using the {\tt JETSET
  7.4} generator.  In the latter case the signal $D$ decays are
removed from the decay table.  Full simulation of the KEDR detector is
performed using the GEANT 3.21 package~\cite{GEANT}.

The likelihood fits yield $M_{D^0}=1865.05\pm 0.33$ MeV and
$M_{D^+}=1869.58\pm 0.49$ MeV.  To obtain the final $D$ masses, one
has to account for a possible deviation of the fit parameters $M_D$
and $\langle \Delta E\rangle$ from the true $D$ masses and energies.
In particular, the central value of $M_D$ can be shifted due to the
asymmetric resolution function and radiative corrections.  The biases
are corrected using the MC simulation.  The final values of the $D$
masses after the corrections are $M_{D^0}=1865.30\pm 0.33$ MeV and
$M_{D^+}=1869.53\pm 0.49$ MeV, where the errors are statistical only.

\vspace{-0.5\baselineskip}
\begin{center}
  \centering
  \scalebox{0.8}{
  \begin{tabular}{|l|c|c|}
    \hline
                                             & $\Delta M_{D^0}$, MeV
                                             & $\Delta M_{D^+}$, MeV\\
    \hline
    Absolute momentum calibration            & 0.04 & 0.04 \\
    Ionization loss in the material          & 0.01 & 0.03 \\
    Momentum resolution                      & 0.13 & 0.10 \\
    ISR corrections                          & 0.16 & 0.11 \\
    Signal PDF                               & 0.07 & 0.05 \\
    Continuum background PDF                 & 0.04 & 0.09 \\
    $D\overline{D}$ background PDF           & 0.03 & 0.06 \\
    Beam energy calibration                  & 0.01 & 0.01 \\
    \hline
    Total                                    & 0.23 & 0.20 \\
    \hline
  \end{tabular}
  }
\vspace{0.5\baselineskip}
\tabcaption{\label{syst} Systematic uncertainties in the $D^0$ and $D^+$ mass measurements.}
\end{center}
\vspace{-1.1\baselineskip}
\begin{center}
  \includegraphics[width=0.88\columnwidth]{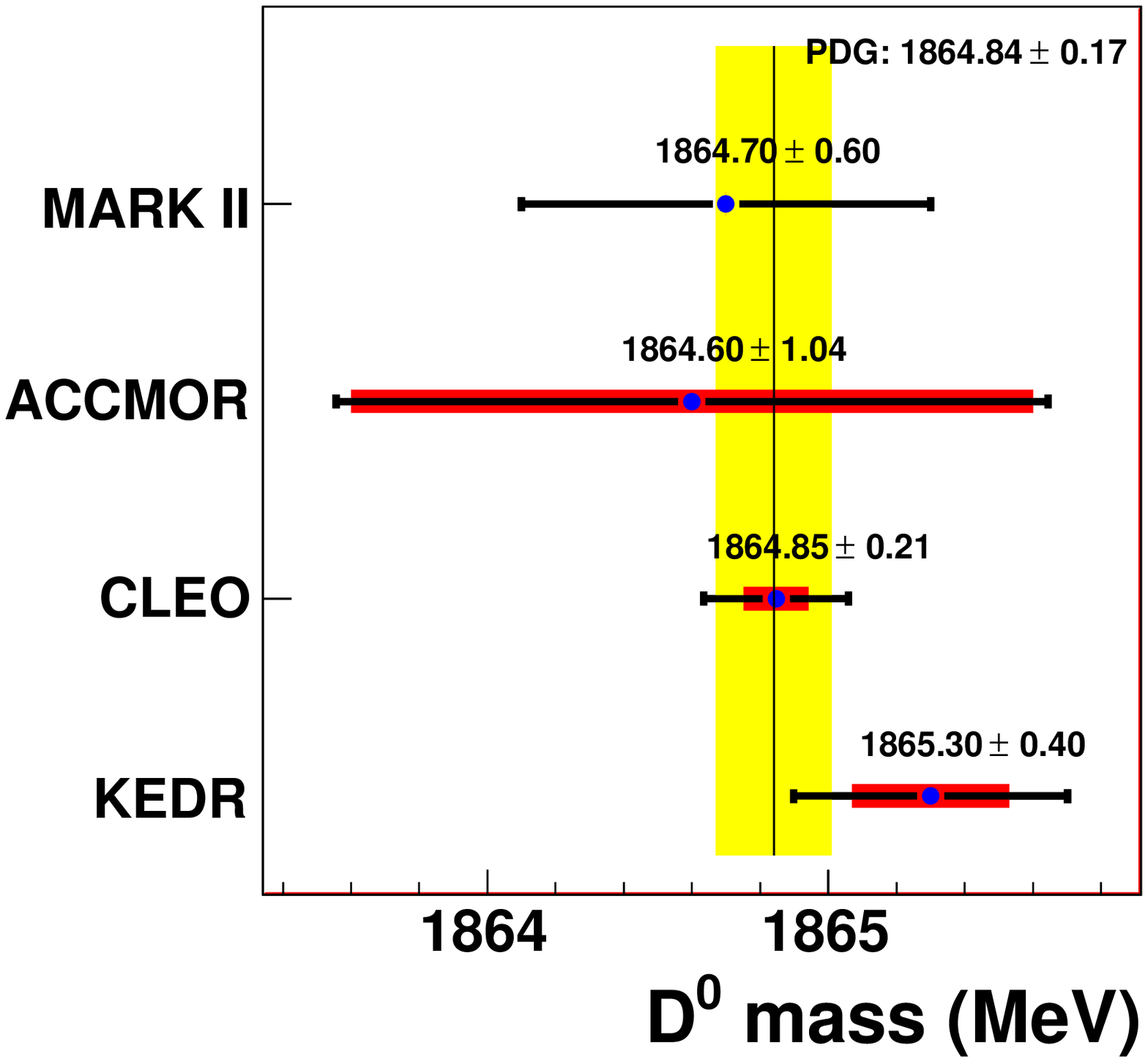}
  \includegraphics[width=0.88\columnwidth]{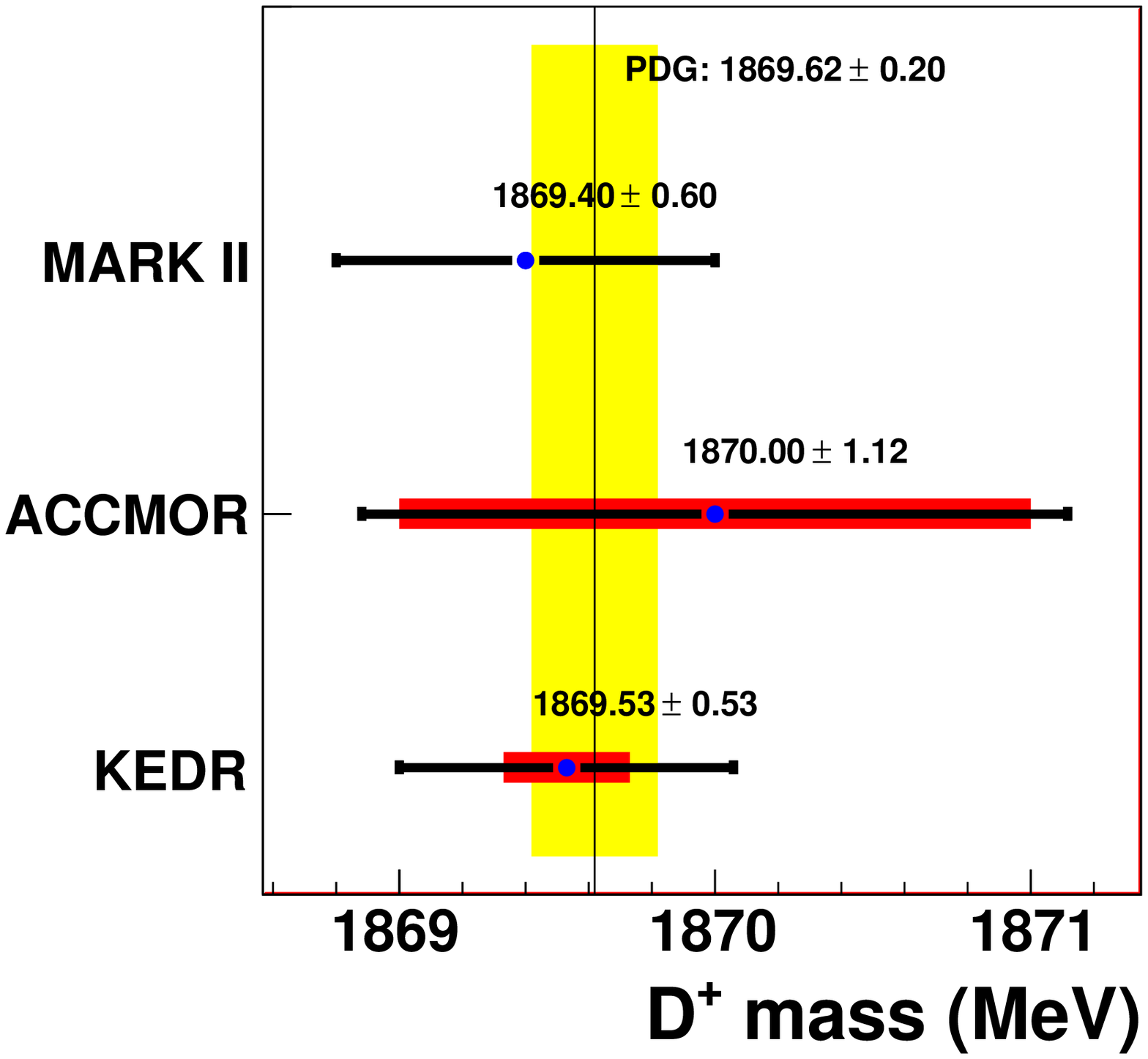}
\vspace{-1.\baselineskip}
  \figcaption{\label{fig:dmass_comp}
     Comparison of KEDR $D$ meson masses with 
          other measurements. The thick and thin error bars show the
          systematic and the total errors, respectively. The shaded
          areas are the PDG-2008 values~\cite{PDG2008}.
          The PDG value for the $D^+$ is obtained using the measured
          mass difference of the $D^+$ and $D^0$ mesons.
             }
\end{center}

Estimates of systematic uncertainties in the $D$ mass measurements are
shown in Table~\ref{syst}.  Figure~\ref{fig:dmass_comp} shows the
comparison of our results with those of the previous experiments.

\section{$\psi(3770)$ mass and widths}
\label{sec:psi3770}

The parameters of the $\psi(3770)$ meson were measured in the scans of
the broad energy region from the \(\psi'\) mass to 3.95 GeV.  The
analysis is based on the luminosity integral of about 2.4 pb$^{-1}$.

Following the similar analyses done
previously\cite{MARKpsi3770,BESpsi3770}, we parameterize tree-level
cross sections of nonresonant $D\bar{D}$ production and $\psi(3770)$
resonance as:
\begin{equation}
  \sigma^{nonres}_{D\bar{D}}(W)=\sigma^{0}(W)+\sigma^{\pm}(W),
\end{equation}
where $\sigma^{0,\pm}(W)=\sigma_{D\bar{D}} \cdot\beta^{3}_{D^{0,\pm}}$, and

\begin{equation}
\sigma_{\psi(3770)}(W) = \frac{3\pi}{M^2} \:\frac{\Gamma_{ee}\Gamma_h}{(W-M)^2+{\Gamma^2(W)}/4},
\end{equation}
where $M$ is the $\psi(3770)$ mass,\\
$\Gamma(W)=\Gamma_{tot}\:
\frac{
  \frac{(R_{0}*P_{D_{0}}(W))^3}{1+(R_{0}*P_{D_{0}}(W))^2}+
  \frac{(R_{0}*P_{D_{\pm}}(W))^3}{1+(R_{0}*P_{D_{\pm}}(W))^2}}
{\frac{(R_{0}*P_{D_{0}}(M))^3}{1+(R_{0}*P_{D_{0}}(M))^2}+
  \frac{(R_{0}*P_{D_{\pm}}(M))^3}{1+(R_{0}*P_{D_{\pm}}(M))^2}}$,\\
$\Gamma_{tot}$ is the $\psi(3770)$ width, and $R_{0}$ is the
interaction radius.

The previous experiments~\cite{MARKpsi3770,BESpsi3770} have ignored
the interference between the $\psi(3770)$ decay and nonresonant
$D\bar{D}$ production.  To investigate its impact on the extracted
parameters of the $\psi(3770)$, we parameterized the total $D\bar{D}$
cross section $\sigma_{D\bar{D}}$ without and with the interference
as:
\begin{equation}
\sigma_{D\bar{D}}(W)= \sigma_{\psi(3770)}(W)+\sigma^{nonres}_{D\bar{D}}(W)
\end{equation}
and
\begin{equation} 
\sigma_{D\bar{D}}(W)= | A_{\psi(3770)}(W) +   A^{nonres}_{D\bar{D}}(W)\cdot e^{i\phi}|^2,
\end{equation} 
respectively, where $|A_{\psi(3770)}(W)|^2 = \sigma_{\psi(3770)}(W)$
$A^{nonres}_{D\bar{D}}(W) = \sqrt{\sigma^{nonres}_{D\bar{D}}(W)}$, and
$\phi$ is the interference phase.

Parameterization of observable cross section is obtained by
convolution of the $\sigma_{D\bar{D}}(W)$ and $\psi'$ cross sections
with the ISR correction function~\cite{KuraevFadin} and beam energy
spread similarly to Eq.~(\ref{equation:convolution}).

Fig.~\ref{fig:3770fits} shows the fits of observable cross section
without and with the interference.  The quality of the latter fit is
significantly better (addition of one fit parameter decreases $\chi^2$
by $\simeq$ 8) and the phase $\phi$ is consistent with $\pi$. The
preliminary results for the $\psi(3770)$ parameters are
\begin{center}
\psfrag{WLabel}{\small{W [MeV]}}
\psfrag{CrossLabel}{$\sigma_{obs}$ [nb] }
\psfrag{ERange}{\hspace{-0.3cm}\small{Fitting region limit}}
\psfrag{Scan2006Label}{Scan 2006}
\psfrag{FitLabel0}{\footnotesize{no interf. ~$\chi^2=17.9/19\,\, {d.o.f.}$} }
\psfrag{FitLabel1}{\footnotesize{with interf. $\chi^2=9.9/18\,\, {d.o.f.}$}}
\includegraphics[width=1.2\columnwidth]{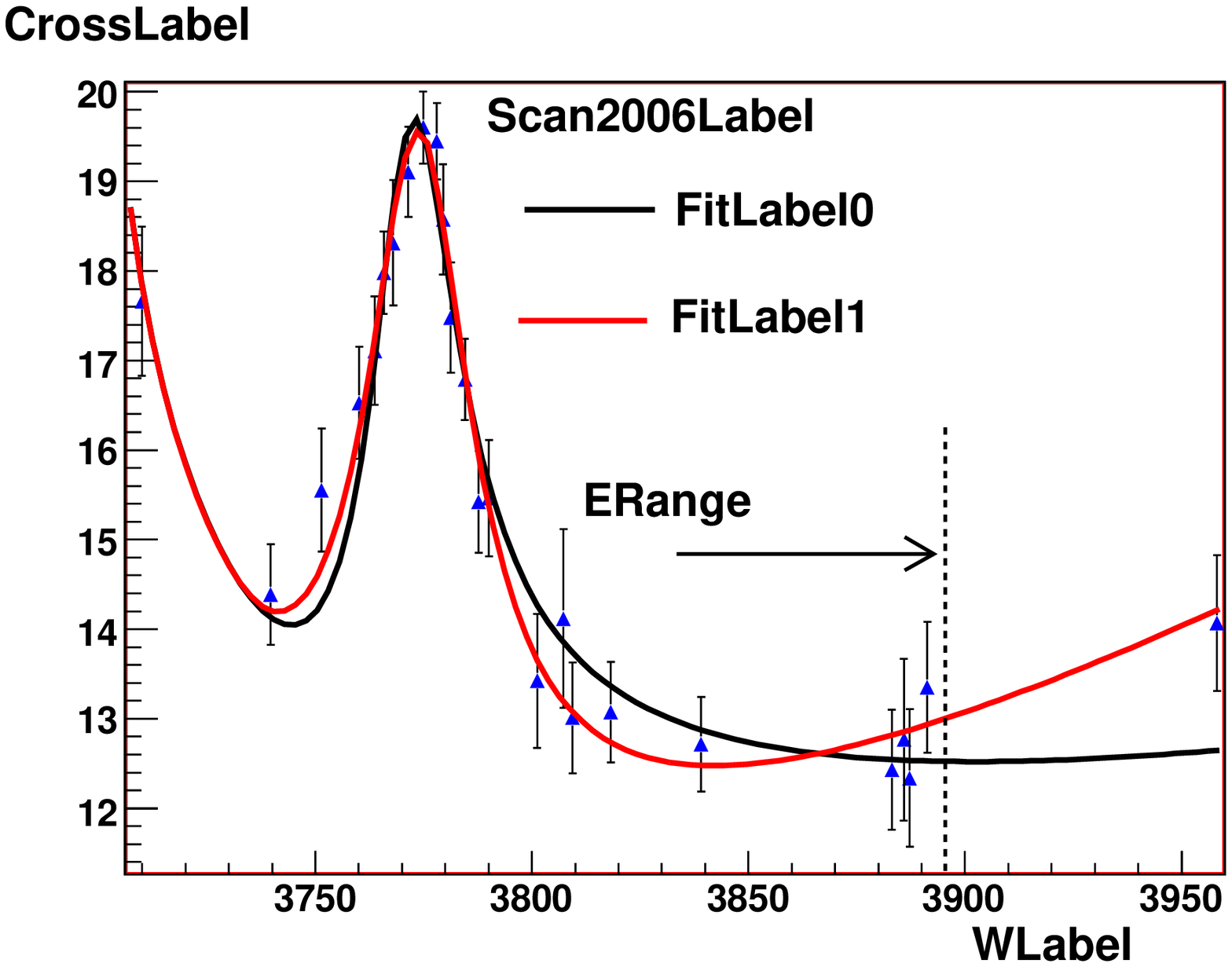}
\figcaption{\label{fig:3770fits} Fits of observable cross section in
  the $\psi(3770)$ region: without and with the interference.}
\end{center}
\begin{itemize*}
 \item $M=3773.2\pm 0.5 \pm 0.6$ MeV,
 \item $\Gamma_{tot}=23.9\pm 2.2 \pm 1.1$ MeV. 
 \item $\Gamma_{ee}=294\pm 22 \pm 30$ eV. 
\end{itemize*}
for the fit without interference, and
\begin{itemize*}
 \item $M=3777.8\pm 1.1 \pm 0.7$ MeV,
 \item $\Gamma_{tot}=28.2\pm 3.1 \pm 2.4$ MeV. 
 \item $\Gamma_{ee}=312\pm 31 \pm 30$ eV.
\end{itemize*}
for the fit with interference. 
Significant increase of $\psi(3770)$ mass in the latter case is apparent.

\begin{center}
\psfrag{KEDRPR}{{~~~~~~~~KEDR}}
\psfrag{PDGUSE}{\tiny{PDG did not use for average}}
\psfrag{DPMASSLABEL}{$M_{\psi{3770}}$[MeV]}
\includegraphics[width=1.\columnwidth]{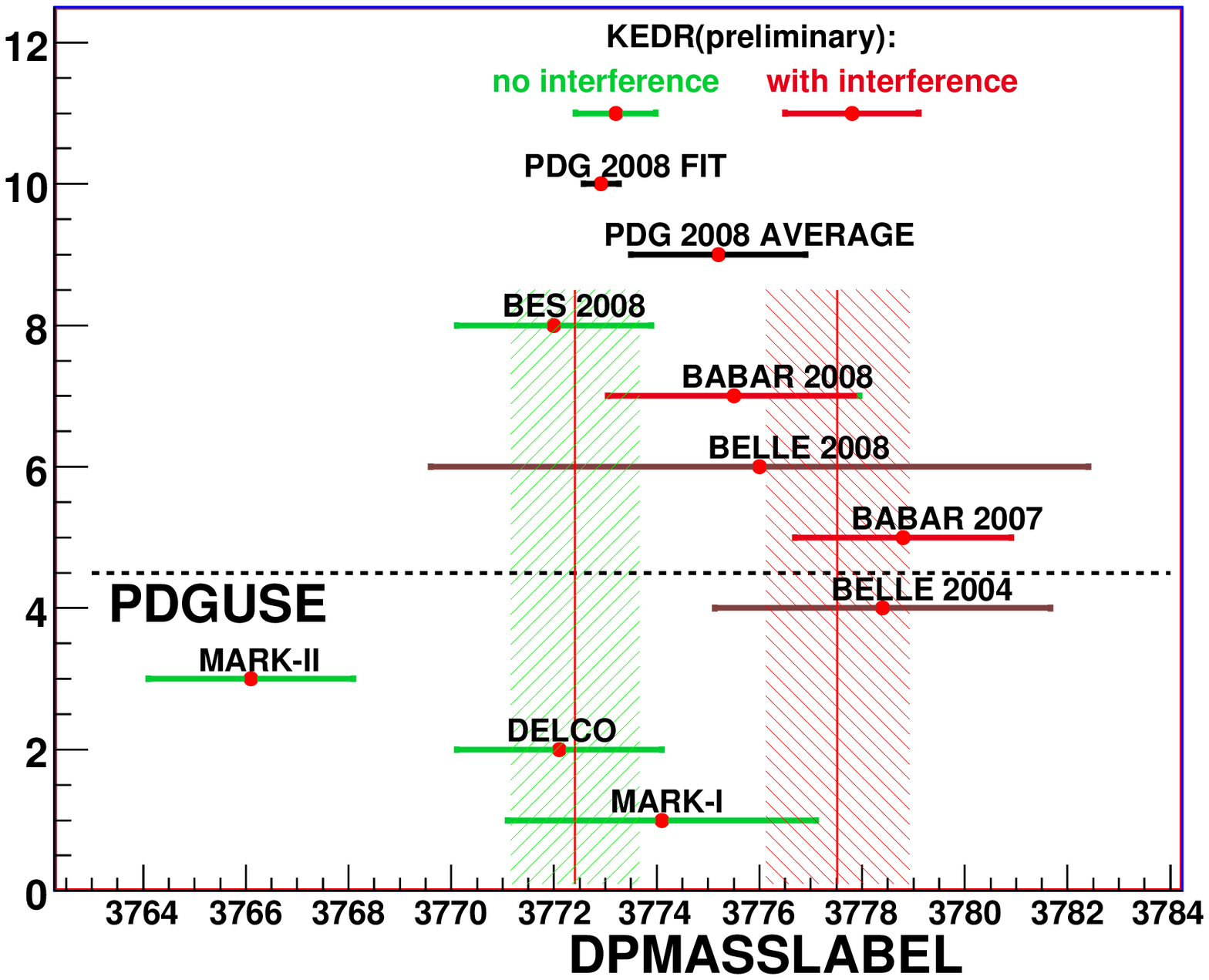}
  \figcaption{\label{fig:compare3770}
Comparison of the $\psi(3770)$ mass measured by KEDR with other experiments.}
\end{center}

Figure~\ref{fig:compare3770} shows comparison of KEDR $\psi(3770)$
masses with results of other experiments.  Mass from our
non-interference fit is consistent with other measurements, which
ignore the interference, while mass from the interference fit is
consistent with the BaBar results~\cite{BABAR2007} which also takes
the interference into account.

\section{Search for narrow resonances}
\label{sec:resonances}

After the \(J/\psi\) discovery, search for other narrow resonances was
performed by several experiments.  The energy region between
\(J/\psi\) and $\Upsilon$ mesons was explored by MARK-1 at SPEAR~\cite{siegrist-1982}, LENA at DORIS~\cite{nickzuporuk-1982}, and
MD-1 at VEPP-4~\cite{blinov-1991}.  The upper limit on the leptonic
width of narrow resonances obtained in these analyses varies between
15 and 970 eV depending on energy.  The search in the energy region
below \(J/\psi\) mass and down to 1.91 GeV was performed only in
experiments at ADONE~\cite{baldini-1975,baldini-1976} with the upper
limit about 500 eV.  Recently KEDR collaboration has revisited the
latter region in view of a recent discovery of unexpected exotic
states above the charm threshold, including the narrow $X(3872)$
state, which proved that surprises are still possible even at low
energies.

\subsection{Experiment description}
The experiment was performed in the beginning of 2009 and the results
are very preliminary.  The energy scan started just above \(J/\psi\)
and finished at 1.85 GeV.  The search for narrow resonances was
conducted by automatic decrease of the center-of-mass energy by about
$2\sigma_w$ (1.4 to 1.9 MeV) steps after collection of required
integrated luminosity in each point.  The integrated luminosity per
energy point varied from 0.3~nb\(^{-1}\) in the upper part of the
energy range to 0.12~nb\(^{-1}\) in the lower one.  The data taken at
each energy were analyzed in real time.  In order to improve
sensitivity, the integrated luminosity was doubled in the energy
points with significant excess of candidate events.

The total integrated luminosity of $\simeq$ 0.28~pb$^{-1}$ was
collected.  The luminosity for data taking was monitored using the
process of single bremsstrahlung, while the analysis uses the offline
measurement based on elastic $e^+e^-$ scattering in the endcap
calorimeter.  The beam energy was measured by the Compton
backscattering technique described in Sec.~\ref{sec:VEPPKEDR}

\subsection{Data analysis}
\label{sec:Data}

The analysis employs three sets of event selection criteria with the
efficiency to hadronic \(J/\psi\) decays ($\epsilon_h$) varying from
44\% to 65\%.  The $\epsilon_h$ was measured from the observable
\(J/\psi\) cross section and known $\Gamma^{J/\psi}_{ee}\cdot
Br(J/\psi \to hadr)$ .  Below we describe the softest set of event
selection criteria.  At the first stage we defined track-level
criteria to select good charged tracks:
\begin{enumerate*}
\item Distances of closest approach to the beam in the transverse
  plane and along the beam axis are less than 0.5 and 10 cm,
  respectively;
\item The deposited energy in the barrel LKr calorimeter is above 20 MeV.
\end{enumerate*}

Then an event-level selection, which account a calorimeter objects
too, was applied:
\begin{enumerate*}
\item The total deposited energy in the calorimeter is above 400 MeV;
\item At least two charged tracks in the event satisfy first criteria
  of the track-level selection;
\item At least one ``good'' charged track satisfies both track-level criteria;
\item There is a charged track acoplanar to the ``good'' one:
  $|\Delta\phi-\pi|>$0.15~rad;
\item Aplanarity of event (sum of momenta transverse to ``event
  plane'') is above $0.1\cdot E_{beam}$
\item There are less than 4 hits in the Muon Chambers;
\item $|\Sigma p^z_i/\Sigma E_i|<0.5$.
\end{enumerate*}

Condition 4 rejects cosmic rays, Bhabhas, and dimuon events. 
Condition 5 rejects radiative Bhabhas and dimuons. 
Condition 6 rejects cosmic ray showers.
Condition 7 rejects two-photon processes and hard ISR.

\subsection{Fit results}
\label{sec:Fit}

The event yield depending on energy is fitted by a function that
assumes the existence of a resonance with mass $M_R$, the leptonic
width $\Gamma^R_{ee}$, and selection efficiency $\epsilon_h$ on top of
flat nonresonant background.  The following parameterization of the
observable cross section was used:
\begin{equation}
\begin{split}
\label{equation:convolution}
\sigma(W)=\sigma_0+\epsilon_{h} \int\! dW'\,&dx \!~\sigma_{R \to hadr}(W')\cdot \\
\cdot& \mathscr{F}(x,W') G\Big(\frac{W-W'}{\sigma_{W}}\Big),
\end{split}
\end{equation}
where
\begin{eqnarray} 
& \sigma_{R \to hadr}(W)=\frac{6\pi^2}{M^2_R}\Gamma^R_{ee}\cdot Br(R\to hadr)\cdot\delta(W-M_R),  \nonumber
\end{eqnarray} 
$\sigma_0$ is the nonresonant background cross section,
$\mathscr{F}(x,W)$ is the radiative correction
function~\cite{KuraevFadin}, $G(x)$ is the Gaussian function.
 
The likelihood fits are performed with the $M_R$ varied in 0.1~MeV
steps with $\Gamma^R_{ee}\cdot Br(R\to hadr)$ as a free parameter, the
energy range in the fit for each $M_R$ value is $M_R\pm 13$ MeV.  The
upper limit obtained with this procedure is $\Gamma^R_{ee}\cdot
Br(R\to hadr)<100$ eV at 90\% confidence level.  In order to get a
more conservative upper limit we took into account that:
\vspace{0.5\baselineskip}
  \begin{itemize*}
  \item $\epsilon_{J/\psi \to hadr}/\epsilon_{e^+ e^- \to hadr}$(3.1
    GeV) $\simeq$ 1.15,
  \item $\epsilon_{e^+ e^- \to hadr}$(3.1 GeV)/$\epsilon_{e^+ e^- \to
      hadr}$(1.9 GeV)$\simeq$1.2,
    \item variation of $\sigma_{W}$ could further increase the limit.
  \end{itemize*}
  \vspace{0.5\baselineskip}
Combining all the factors, we set a limit \\
$\Gamma^R_{ee}\cdot Br(R\to hadr) < 150$ eV (90 \% {\em c.l.}) \\
in the W range between 1.85 GeV and $M_{J/\Psi}$.

\section{Results}

Masses of charged and neutral $D$ mesons are obtained at KEDR experiment: 
\begin{itemize*}
 \item $M_{D^0}=1865.30\pm 0.33\pm 0.23$ MeV,
 \item $M_{D^+}=1869.53\pm 0.49\pm 0.20$ MeV. 
\end{itemize*}
The $D^0$ mass value is consistent with the more precise measurement
of the CLEO collaboration~\cite{cleo}, while that of the $D^+$ mass is
presently the most precise direct determination.

The preliminary \(\psi(3770)\) parameters are obtained with standard
non-interference parameterization~\cite{MARKpsi3770,BESpsi3770}:
\begin{itemize*}
 \item $M=3773.2\pm 0.5 \pm 0.6$ MeV,
 \item $\Gamma_{tot}=23.9\pm 2.2 \pm 1.1$ MeV,
 \item $\Gamma_{ee}=294\pm 22 \pm 30$ eV. 
\end{itemize*}
The parameters with the interference:
\begin{itemize*}
 \item $M=3777.8\pm 1.1 \pm 0.7$ MeV,
 \item $\Gamma_{tot}=28.2\pm 3.1 \pm 2.4$ MeV, 
 \item $\Gamma_{ee}=312\pm 31 \pm 30$ eV. 
\end{itemize*}
Taking the interference into account significantly improves the fit
quality and increases the $\psi(3770)$ mass by 4.6 MeV.

A preliminary upper limit for narrow resonances in the W range between
1.85 GeV and $M_{J/\Psi}$:
\begin{itemize*}
\item $\Gamma^R_{ee}\cdot Br(R\to hadr) < 150$ eV (90 \% {\em c.l.}).
\end{itemize*}

\end{multicols}

\vspace{-2mm}
\centerline{\rule{80mm}{0.1pt}}
\vspace{2mm}

\begin{multicols}{2}

\bibliographystyle{h-physrev5}
\bibliography{proc}

\end{multicols}

\end{document}